\documentclass[aps,prl,twocolumn,showpacs,superscriptaddress,groupedaddress,nofootinbib]{revtex4-2}  
\usepackage{cancel}
\usepackage{graphicx} 
\usepackage[T1]{fontenc}
\usepackage{lmodern} 
\usepackage{graphicx}  
\usepackage{dcolumn}   
\usepackage{bm}        
\usepackage{amssymb}   
\usepackage{amsmath}
\usepackage{bbold}
\usepackage{array}
\usepackage{float}
\usepackage{makecell}
\usepackage{braket}
\usepackage{longtable}
\usepackage{supertabular,booktabs}
\usepackage{soul}

\usepackage{titlesec} 
\usepackage[colorlinks,citecolor=blue,urlcolor=blue,hypertexnames=true]{hyperref}
\usepackage{subfigure}

\usepackage[usenames,dvipsnames,svgnames,table]{xcolor} 

\renewcommand{\arraystretch}{2}

\newcommand{\be}{\begin{equation}}
\newcommand{\ee}{\end{equation}}
\newcommand{\bea}{\begin{eqnarray}}
\newcommand{\eea}{\end{eqnarray}}
\newcommand{\bml}{\begin{subequations}}
\newcommand{\eml}{\end{subequations}}
\newcommand{\bfig}{\begin{figure}}
\newcommand{\efig}{\end{figure}}

\newcommand{\slashed}{\hspace{-1.1ex}/}

\newcommand{\bmat}{\begin{pmatrix}}
\newcommand{\emat}{\end{pmatrix}}
\usepackage{graphicx, slashed,booktabs, color, multirow, float,
amsfonts, bbold, mathtools, sidecap, tikz, bm,enumitem}
\usepackage{multirow}
\usepackage{bbding}
\usepackage{titlesec}
\usepackage{hyperref}
\usepackage{wasysym}
\usepackage{amssymb}
\usepackage{pifont}
\newcommand{\grad}{\nabla}

\renewcommand{\leq}{\leqslant}

\usepackage[dvipsnames, usenames]{xcolor}

\definecolor{linkcolor}{rgb}{0.55, 0.13, .32}

\definecolor{oucrimsonred}{rgb}{0.6, 0.0, 0.0}
\definecolor{persianblue}{rgb}{0.11, 0.22, 0.73}
\definecolor{forestgreen}{rgb}{0.13,0.35,0.13}
\definecolor{lightgray}{rgb}{0.83, 0.83, 0.83}
 \hypersetup{colorlinks, citecolor=oucrimsonred, linkcolor=persianblue, urlcolor=oucrimsonred}
\definecolor{cornellred}{rgb}{0.7, 0.11, 0.11}
\definecolor{navyblue}{rgb}{0.0, 0.0, 0.5}
\definecolor{amethyst}{rgb}{0.6, 0.4, 0.8}
\definecolor{yellow}{rgb}{1.0, 1.0, 0.0}
\definecolor{firebrick}{rgb}{0.7, 0.13, 0.13}
\definecolor{tangerineyellow}{rgb}{1.0, 0.8, 0.0}
\definecolor{deepfuchsia}{rgb}{0.76, 0.33, 0.76}
\definecolor{amber}{rgb}{1.0, 0.75, 0.0}
\definecolor{VioletRed4}{rgb}{0.55, 0.13, .32}
\definecolor{indiagreen}{rgb}{0.07, 0.53, 0.03}
\definecolor{VioletRed4}{rgb}{0.55, 0.13, .32}

\usepackage{hyperref}
\usepackage{graphics, appendix, afterpage, makecell} 
\usepackage{bbold}
\usepackage{tikz}
\usepackage{adjustbox}

\usepackage{tcolorbox}

\definecolor{oucrimsonred}{rgb}{0.6, 0.0, 0.0}
\definecolor{persianblue}{rgb}{0.11, 0.22, 0.73}
\definecolor{forestgreen}{rgb}{0.13,0.35,0.13}
\definecolor{lightgray}{rgb}{0.83, 0.83, 0.83}
 \hypersetup{colorlinks, citecolor=oucrimsonred, linkcolor=persianblue, urlcolor=oucrimsonred}
\definecolor{cornellred}{rgb}{0.7, 0.11, 0.11}
\definecolor{navyblue}{rgb}{0.0, 0.0, 0.5}
\definecolor{amethyst}{rgb}{0.6, 0.4, 0.8}
\definecolor{yellow}{rgb}{1.0, 1.0, 0.0}
\definecolor{firebrick}{rgb}{0.7, 0.13, 0.13}
\definecolor{tangerineyellow}{rgb}{1.0, 0.8, 0.0}
\definecolor{deepfuchsia}{rgb}{0.76, 0.33, 0.76}
\definecolor{amber}{rgb}{1.0, 0.75, 0.0}
\definecolor{VioletRed4}{rgb}{0.55, 0.13, .32}
\definecolor{indiagreen}{rgb}{0.07, 0.53, 0.03}
\definecolor{VioletRed4}{rgb}{0.55, 0.13, .32}

\definecolor{oucrimsonred}{rgb}{0.6, 0.0, 0.0}
\newcommand\vertarrowbox[3][6ex]{%
  \begin{array}[t]{@{}c@{}} #2 \\
  \left\uparrow\vcenter{\hrule height #1}\right.\kern-\nulldelimiterspace\\
  \makebox[0pt]{\scriptsize#3}
  \end{array}%
}

\definecolor{mtcolor}{rgb}{.8,.3,.1}

\definecolor{violachiaro}{rgb}{1,0.6,1}

\definecolor{gbcolor}{rgb}{.43,.22,.12}
 
\definecolor{gbcolor2}{rgb}{.9,.2,.6}
\definecolor{gbcolor3}{rgb}{.3,.2,.6}

\definecolor{verdechiaro}{rgb}{0.6,1,0.6}
\definecolor{giallochiaro}{rgb}{1,1,0.6}
\definecolor{bluscuro}{rgb}{0.15, 0.2, 0.9}
\definecolor{verdes}{rgb}{0.1, 0.5, 0.1}%
\definecolor{tangerineyellow}{rgb}{1.0, 0.8, 0.0}
\definecolor{smokyblack}{rgb}{0.06, 0.05, 0.03}

\definecolor{americanrose}{rgb}{1.0, 0.01, 0.24}
\definecolor{cobalt}{rgb}{0.0, 0.28, 0.67}
\definecolor{brandeisblue}{rgb}{0.0, 0.44, 1.0}
\definecolor{mycolor}{rgb}{0.0, 0.0, 0.5}
\definecolor{oxfordblue}{rgb}{0.0, 0.13, 0.28}
\definecolor{azure}{rgb}{0.0, 0.5, 1.0}
\definecolor{turquoiseblue}{rgb}{0.0, 1.0, 0.94}
\newtcolorbox{mynewbox}[1]{colback=white!5!white,colframe=azure!75!black,fonttitle=\bfseries,title=#1}
\newtcolorbox{mybox}{colback=mycolor!5!white,colframe=azure!75!black}
\newtcolorbox{mynamedbox}[1]{colback=mycolor!5!white,colframe=azure!75!black,title=#1}
\definecolor{venetianred}{rgb}{0.78, 0.03, 0.08}
\newtcolorbox{mynamedbox1}[1]{colback=venetianred!5!white,colframe=venetianred!80!black,title=#1}
\newtcolorbox{mynamedbox2}[1]{colback=azure!5!white,colframe=azure!80!black,title=#1}

\definecolor{rossocorsa}{rgb}{0.83, 0.0, 0.0}

\tikzset{->-/.style={decoration={
  markings,
  mark=at position #1 with {\arrow{>}}},postaction={decorate}}}
\tikzset{-<-/.style={decoration={
  markings,
  mark=at position #1 with {\arrow{<}}},postaction={decorate}}} 

\def\be{\begin{equation}}
\def\ee{\end{equation}}
\def\ba{\begin{eqnarray}}
\def\ea{\end{eqnarray}}

\def\L*{{\cal L}_*}
\def\L{\mathcal{L}}
\def\({\left(}
\def\){\right)}

\def\<{\langle}
\def\>{\rangle}

\def\cs2{c_{s}^{2}}

 \def\be   {\begin{equation}}   \def\ee   {\end{equation}}
 \def\ba   {\begin{array}}      \def\ea   {\end{array}}
 \def\bea  {\begin{eqnarray}}   \def\eea  {\end{eqnarray}}
 \def\bean {\begin{eqnarray*}}  \def\eean {\end{eqnarray*}}

\titleclass{\subsubsubsection}{straight}[\subsection]

\newcounter{subsubsubsection}[subsubsection]
\renewcommand\thesubsubsubsection{\thesubsubsection.\arabic{subsubsubsection}}

\titleformat{\subsubsubsection}
  {\normalfont\normalsize\bfseries}{\thesubsubsubsection}{1em}{}
\titlespacing*{\subsubsubsection}
{0pt}{3.25ex plus 1ex minus .2ex}{1.5ex plus .2ex}

\makeatletter
\renewcommand\paragraph{\@startsection{paragraph}{5}{\z@}%
  {3.25ex \@plus1ex \@minus.2ex}%
  {-1em}%
  {\normalfont\normalsize\bfseries}}
\renewcommand\subparagraph{\@startsection{subparagraph}{6}{\parindent}%
  {3.25ex \@plus1ex \@minus .2ex}%
  {-1em}%
  {\normalfont\normalsize\bfseries}}
\def\toclevel@subsubsubsection{4}
\def\toclevel@paragraph{5}
\def\toclevel@paragraph{6}
\def\l@subsubsubsection{\@dottedtocline{4}{7em}{4em}}
\def\l@paragraph{\@dottedtocline{5}{10em}{5em}}
\def\l@subparagraph{\@dottedtocline{6}{14em}{6em}}
\makeatother

\usepackage{titlesec} 
\usepackage[colorlinks,citecolor=blue,urlcolor=blue,hypertexnames=true]{hyperref}
\usepackage{subfigure}

\usepackage[usenames,dvipsnames,svgnames,table]{xcolor}

\usepackage{tikzsymbols}
\usepackage{natbib}
\usepackage{float}

\usepackage{tikz,xcolor,hyperref}

\usepackage{slashed}

\definecolor{lime}{HTML}{A6CE39}
\DeclareRobustCommand{\orcidicon}{
	\begin{tikzpicture}
	\draw[lime, fill=lime] (0,0) 
	circle [radius=0.2] 
	node[white] {{\fontfamily{qag}\selectfont \tiny ID}};
	\draw[white, fill=white] (-0.0625,0.095) 
	circle [radius=0.007];
	\end{tikzpicture}
	\hspace{-2mm}
}

\foreach \x in {A, ..., Z}{\expandafter\xdef\csname orcid\x\endcsname{\noexpand\href{https://orcid.org/\csname orcidauthor\x\endcsname}
			{\noexpand\orcidicon}}
}

\usepackage{bbold}
\usepackage{tikz}
\usepackage{adjustbox}
\usepackage{tcolorbox}
\usepackage{enumitem}
\usepackage{amsfonts}

\setlist[itemize,1]{label=$\times$}
\setlist[itemize,2]{label=$\checkmark$}
\setlist[itemize,3]{label=$\diamond$}
\setlist[itemize,4]{label=$\bullet$}

\begin{document}
\title{\Large \textcolor{Sepia}{Evading no-go for PBH formation and production of SIGWs using Multiple Sharp Transitions in EFT of single field inflation}}
\author{\large  Gourab Bhattacharya\orcidA{}${}^{1,2,3}$}
\email{gourabphy123@gmail.com}
\author{\large Sayantan Choudhury\orcidB{}${}^{1}$}
\email{sayantan\_ccsp@sgtuniversity.org,  \\ sayanphysicsisi@gmail.com (Corresponding author) }
\author{\large Kritartha Dey\orcidC{}${}^{1}$}
\email{kritartha09@gmail.com }
\author{ \large Saptarshi Ghosh\orcidD{}${}^{1,4}$}
\email{saptarshighosh422@@gmail.com}
\author{\large Ahaskar Karde\orcidE{}${}^{1}$}
\email{kardeahaskar@gmail.com}
\author{ \large Navneet Suryaprakash Mishra\orcidF{}${}^{1}$}
\email{nmts1309@gmail.com}
\affiliation{ ${}^{1}$Centre For Cosmology and Science Popularization (CCSP),\\
        SGT University, Gurugram, Delhi- NCR, Haryana- 122505, India,}
\affiliation{${}^{2}$School of Physical Sciences,  National Institute of Science Education and Research, Bhubaneswar, Odisha - 752050, India,}
\affiliation{${}^{3}$ Homi Bhabha National Institute, Training School Complex, Anushakti Nagar, Mumbai - 400085, India,}
\affiliation{${}^{4}$ Department of Physics, Ramakrishna Mission Residential College (Autonomous), Narendrapur, Kolkata 700103}

\begin{abstract}

  Deploying \textit{multiple sharp transitions} (MSTs) under a unified framework, we investigate the formation of Primordial Black Holes (PBHs) and the production of Scalar Induced Gravitational Waves (SIGWs) by incorporating one-loop corrected renormalized-resummed scalar power spectrum. With effective sound speed parameter, $1 \leq c_s \leq 1.17$, the direct consequence is the generation of PBH masses spanning  $M_{\rm PBH}\sim{\cal O}(10^{-31}M_{\odot}- 10^{4}M_{\odot})$,  thus evading well known \textit{No-go theorem} on PBH mass. Our results align coherently with the extensive NANOGrav 15-year data and the sensitivities outlined by other terrestrial and space-based experiments (e.g.: LISA, HLVK, BBO, HLV(O3), etc.).
\end{abstract}

\maketitle

\newpage
Numerous recent developments have brought attention to the study of Primordial Black Hole (PBH) formation \cite{Hawking:1974rv,Carr:1974nx,Carr:1975qj,Chapline:1975ojl,Carr:1993aq,Choudhury:2013woa} from the inflationary paradigm. Among these are the canonical and non-canonical scalar field models, some of which involve smooth or sharp transitions from a slow-roll (SR) into an ultra-slow roll (USR) phase \cite{Kristiano:2022maq,Riotto:2023hoz,Choudhury:2023vuj,Choudhury:2023jlt,Choudhury:2023rks,Choudhury:2023hvf,Choudhury:2023kdb,Choudhury:2023hfm,Kristiano:2023scm,Riotto:2023gpm,Firouzjahi:2023ahg,Firouzjahi:2023aum,Franciolini:2023lgy,Cheng:2023ikq,Tasinato:2023ukp,Motohashi:2023syh, Banerjee:2021lqu}. Within the framework of Effective Field Theory (EFT) for single field inflation \cite{Weinberg:2008hq,Cheung:2007st,Choudhury:2017glj}, there have been attempts to accommodate the diverse physics of these models, operating under an effective action that remains valid beneath a specific UV cut-off scale. We choose to conduct our analysis within the territory of EFT.

One of the enduring challenges has been generating PBHs with sufficiently large masses, ${\cal O}(M_{\odot})$ (solar mass) while considering quantum loop corrections within the scalar power spectrum. Earlier works with single-field inflation models in the EFT setup have discussed this problem in great detail \cite{Choudhury:2023vuj,Choudhury:2023jlt,Choudhury:2023rks}. Yet, a key limitation has been the absence of renormalization and resummation procedures in these approaches, which are crucial for a robust inflationary framework.
Inflation featuring multiple sharp transitions (MSTs) introduces a distinctive approach to resolving the ongoing PBH formation challenge comprehensively. Our analysis will showcase how MSTs provide the flexibility to cover the entire wavenumber spectrum required for PBH production, spanning $M_{\rm PBH}\sim{\cal O}(10^{-31}M_{\odot}-10^{4}M_{\odot})$. This achievement evades the previously established \textit{No-go theorem} regarding PBH mass constraints \cite{Choudhury:2023vuj,Choudhury:2023jlt,Choudhury:2023rks}. We will demonstrate that the conditions for successful inflation are met even in the presence of a one-loop corrected renormalized and resummed scalar power spectrum (OLRRSPS) with MSTs.
Our investigation revolves around the characteristics of the OLRRSPS related to the PBH formation mechanism and producing the spectrum of scalar-induced gravitational waves (SIGWs) across a broad frequency range corresponding to the aforementioned PBH masses. The resulting SIGW spectrum is poised to explain the NANOGrav 15-year signal \cite{NANOGrav:2023hvm} and remains consistent with the sensitivities of existing and proposed terrestrial and space-based gravitational wave experiments, including LISA \cite{LISA:2017pwj}, DECIGO \cite{Kawamura:2011zz}, ET \cite{Punturo:2010zz}, CE \cite{Reitze:2019iox}, BBO \cite{Crowder:2005nr}, HLVK \cite{LIGOScientific:2014pky, VIRGO:2014yos, KAGRA:2018plz}, and HLV(O3). These compelling aspects of MSTs depict a promising scenario deserving of thorough investigation, encapsulating the primary aim of this letter.

To this effect, we begin with the EFT setup for inflation which is described by the following representative action \cite{Cheung:2007st,Choudhury:2017glj}:
\begin{widetext}
    \bea \label{action}
  S &=& \int d^4 x \sqrt{-g} \Bigg[\frac{M_p^2}{2}R+M_{p}^2 \dot H g^{00}-M_p ^2 (3H^2 +\dot H)+\sum_{n=2,3}\frac{M_{n}^{4}(t)}{n!}(g^{00}+1)^{n}-\frac{\overline{M}_{1}^3 (t)}{2} (g^{00}+1) \delta K_{\mu}^{\mu} \nonumber\\ 
&&\quad\quad\quad\quad\quad\quad\quad\quad\quad\quad\quad\quad\quad\quad\quad\quad\quad\quad\quad\quad\quad\quad\quad\quad\quad\quad\quad\quad-\frac{\overline{M}_{2}^2 (t)}{2} (\delta K_{\mu}^{\mu})^2 - \frac{\overline{M}_{3}^2 (t)}{2} \delta K_{\nu}^{\mu}\delta K_{\mu}^{\nu}+\cdots\Bigg]. 
    \quad\eea    
\end{widetext}
Here the scalar perturbation, $\delta\phi$, has the following non-linear transformation under time diffeomorphisms: $\quad t \rightarrow t + \xi^{0}(t,\textbf{x}),\;x^{i} \rightarrow x^{i}$, as $\delta\phi \rightarrow \delta\phi + \dot{\phi_{0}}(t)\xi^{0}(t,\textbf {x}),\;\forall i=1,2,3$, where $\xi^{0}(t,\bf x)$ is the time diffeomorphism parameter and $\phi_{0}(t)$ represents the background scalar field embedded in the spatially flat FLRW space-time.
Using the gravitational gauge, $\phi(t,\textbf x) = \phi_{0}(t)$, one can write \bea\delta\phi(t,\textbf{x}) = 0.\eea
Various terms in the action are: $g^{00}$, the temporal part of the background metric, $M_p$ denotes the reduced Planck Mass, $H$ represents the Hubble parameter in quasi de-Sitter space, and $K_{\mu \nu}$ is the extrinsic curvature. The coefficients
$M_{2}(t)$, $M_{3}(t)$, $\overline{M}_{1}(t)$, $\overline{M}_{2}(t)$, $\overline{M}_{3}(t)$ represent the slowly-varying time-dependent Wilson coefficients for this EFT. This action exhibits broken time diffeomorphisms, but introducing a new scalar field $\pi(t,\textbf{x})$ helps to restore the broken symmetry.  This is the  Goldstone mode which transforms non-linearly under time diffeomorphism, 
$\tilde{\pi}(t,\textbf{x}) = -\xi^{0}(t,\textbf{x})+\pi(t,\textbf{x})$.
Here, $\pi(t,\textbf{x})$ represents the scalar perturbations around the background and $ \tilde{\pi}(t,\textbf{x})$ represents the shifted version of the Goldstone field. The condition $\pi(t,\textbf{x}) = 0$ promotes the parameter $\xi^{0}(t,\textbf{x})$ to the field $\tilde{\pi}(t,\textbf{x})$. This is the St\"{u}ckelberg technique which is a method similar to spontaneous symmetry breaking in $SU(N)$ non-abelian gauge theories. The terms such as $M_{p}^{2}\dot{H}\dot{\pi}\delta g^{00}$ are neglected above the energy scale $E_{\rm mix} = \sqrt{\dot{H}}$ called the decoupling limit. While terms such as $-M_{p}^{2}\dot{H}\dot{\pi}^{2}$ survive and later contribute towards the building of the action in terms of the Goldstone modes. For additional information on this aspect see ref.\cite{Cheung:2007st,Choudhury:2017glj}. 

The Goldstone mode has a one-to-one mapping in the linear regime with the comoving curvature perturbation through the relation, $\zeta (t,\textbf{x}) = - {H}{\pi}(t,\textbf{x})$ \cite{Cheung:2007st,Choudhury:2017glj}, in terms of which the second-order action in the conformal time coordinates can be written as:
\bea
    \label{s3pi}
     S_{\zeta}^{(2)} &=& M_p ^2 \int d\tau \; d^3x\; a^2 \frac{\epsilon}{c_s ^2} \Big [\zeta'^2-c_s ^2(\partial_i \zeta )^2 \Big]. 
\eea
where the effective sound is given by: \bea c_s = \left(1 - \frac{2M_2^4}{\dot{H}M_p^2}\right)^{-1/2},\eea and the first SR parameter, $\epsilon$ can be expressed as $\epsilon= \left(1-{\cal H^{\prime}}/{\cal H}^{2}\right)$. We introduce a re-scaled variable $v=zM_p\zeta$ with $z=a\sqrt{2\epsilon}/c_s$ and by varying the action in Fourier transformed space we acquire the Mukhanov-Sasaki (MS) equation:
\bea
  \label{MS}
  v_{\textbf{k}}^{\prime \prime } + \bigg [c_s ^2 k^2 - \frac{z^{\prime \prime}}{z} \bigg]v_{\textbf{k}}=0, \quad \quad {\rm where}\quad\frac{z^{\prime \prime}}{z} \approx \frac{2}{\tau ^2}.
\eea
where $
k=|\textbf{k}|$ refers to the magnitude of the momentum vector $\textbf{k}$.

In our current scenario of MSTs, we begin with an initial slow-roll phase (SR$_{1}$). During the remaining conformal time period till the end of inflation, we encounter multiple USR and SR phases. To address this, we first provide the general solution for the perturbed scalar mode after combining all the phases such as:
\bea \label{solsr1}
&&\zeta_{\textbf{k}}^{\rm SR_{1}}(\tau) =F_{\textbf{k}}\bigg[\alpha_{\textbf{k}}^{(1)}g_{+,\textbf{k}}(\tau)-\beta_{\textbf{k}}^{(1)}g_{-,\textbf{k}}(\tau)\bigg],
\\ \label{solusr}
&&\zeta_{\textbf{k}}^{\rm USR_{n}}(\tau) = F_{\textbf{k}} 
\left(\frac{\tau_{s_n}} {\tau}\right)^{3}\bigg[\alpha_{\textbf{k}}^{(2n)}g_{+,\textbf{k}}(\tau)\nonumber\\
&& \quad\quad\quad\quad\quad\quad\quad\quad\quad\quad\quad\quad\quad\quad -\beta_{\textbf{k}}^{(2n)}g_{-,\textbf{k}}(\tau)\bigg], 
\\ \label{solsrn}
&&\zeta_{\textbf{k}}^{\rm SR_{n+1}}(\tau) = F_{\textbf{k}}\left(\frac{\tau_{s_{n}}} {\tau_{e_n}}\right)^{3} \bigg[\alpha_{\textbf{k}}^{(2n+1)}g_{+,\textbf{k}}(\tau) \nonumber\\
&& \quad\quad\quad\quad\quad\quad\quad\quad\quad\quad\quad\quad\quad-\beta_{\textbf{k}}^{(2n+1)}g_{-,\textbf{k}}(\tau)\bigg],\quad
\quad\eea
where we have defined the following quantities for simplification purposes:
\bea
F_{\textbf{k}} &=& \left(\frac{ic_{s}H}{2M_{\rm p}\sqrt{\epsilon}}\right)_{*}\frac{1}{(c_{s}k)^{3/2}},\\
g_{\pm,\textbf{k}}(\tau) &=& (1\pm ikc_s\tau)\exp{(\mp ikc_s\tau)}.
\eea
In the above expressions, eqn.(\ref{solsr1}) corresponds to the mode solution for the SR$_{1}$ phase, which contains the pivot scale set at the value $k_{*} = 0.02{\rm Mpc}^{-1}$ and the expression in the parenthesis is evaluated at the pivot scale. The quantum initial condition in the SR$_{1}$ phase is fixed in terms of the Bunch Davies initial condition, which is implemented via:
$\alpha_{\textbf{k}}^{(1)}=1,\;\beta_{\textbf{k}}^{(1)}=0$. The eqns.(\ref{solusr}, \ref{solsrn}) correspond to the generalized version of the mode functions in the $n$th phase of the USR (USR$_{n}$) and the $(n+1)$th SR phase (SR$_{n+1}$) respectively. The variable $n=1,2,\cdots$ is used to count the number of sharp transitions in the present scenario and its exact range will be justified shortly from the conditions for successful inflation. After the SR$_{1}$ phase ($\tau < \tau_{s_{1}}$) comes to an end at conformal time $\tau=\tau_{s_{1}}$, we observe a sharp transition into the USR$_{1}$ phase which is described using the conformal time interval $\tau_{s_{1}} \leq \tau \leq \tau_{e_{1}}$. At the conformal time $\tau=\tau_{e_{1}}$, another sharp transition occurs which marks the exit of the USR$_{1}$ phase and also the beginning of the next SR phase SR$_{2}$ which then continues for the time interval $\tau_{e_{1}} \leq \tau \leq \tau_{s_{2}}$. Such MSTs will continue to occur
in the fashion SR$_{1}$ $\longrightarrow$USR$_{1}$ $\longrightarrow$SR$_{2}$ $\longrightarrow$USR$_{2}$ $\longrightarrow$SR$_{3}$ $\longrightarrow \cdots$
until successful inflation is achieved in the present context. The Bogoliubov coefficients after the ${\rm SR_{1}}$ phase are shifted from the Bunch Davies to the Non-Bunch Davies vacuum state. The coefficients contained in the above mode solutions arise as a result of the Israel junction conditions at the boundaries for the sharp transitions implemented at the conformal times $\tau=\tau_{s_{n}}$ and $\tau=\tau_{e_{n}}$. We now directly mention the general form of such coefficients as: 
\begin{widetext}
\bea
\alpha^{(2n)}_{\textbf{k}}&=&-\frac{k_{e_{n-1}}^{3}k_{s_{n}}^{3}}{2k^{3}k_{s_{n-1}}^{3}}
\bigg[\bigg(3i-\frac{2k^{3}}{k_{s_{n}}^{3}}+\frac{3ik^{2}}{k_{s_{n}}^{2}}\bigg)\alpha^{(2n-1)}_{\textbf{k}}+3i\exp{\left(-\frac{2ik}{k_{s_{n}}}\right)}\bigg(i-\frac{k}{k_{s_{n}}}\bigg)^{2}\beta^{(2n-1)}_{\textbf{k}}\bigg],\\
\beta^{(2n)}_{\textbf{k}}&=&
\exp{\left(\frac{2ik}{k_{s_{n}}}\right)}\frac{k_{e_{n-1}}^{3}k_{s_{n}}^{3}}{2k^{3}k_{s_{n-1}}^{3}}
\bigg[3i\bigg(i+\frac{k}{k_{s_{n}}}\bigg)^{2}\alpha^{(2n-1)}_{\textbf{k}}+\exp{\left(-\frac{2ik}{k_{s_{n}}}\right)}\bigg(3i+\frac{2k^{3}}{k_{s_{n}}^{3}}+\frac{3ik^{2}}{k_{s_{n}}^{2}}\bigg)
\beta^{(2n-1)}_{\textbf{k}}\bigg],\\
\alpha^{(2n+1)}_{\textbf{k}}&=&\frac{k_{e_{n}}^{3}}{2k^{3}}
\bigg[\bigg(3i+\frac{2k^{3}}{k_{e_{n}}^{3}}+\frac{3ik^{2}}{k_{e_{n}}^{2}}\bigg)\alpha^{(2n)}_{\textbf{k}}+3i\exp{\left(-\frac{2ik}{k_{e_{n}}}\right)}\bigg(i-\frac{k}{k_{e_{n}}}\bigg)^{2}\beta^{(2n)}_{\textbf{k}}\bigg],\\
\beta^{(2n+1)}_{\textbf{k}} &=& -\exp{\left(\frac{2ik}{k_{e_{n}}}\right)}\frac{k_{e_{n}}^{3}}{2k^{3}}
\bigg[3i\bigg(i+\frac{k}{k_{e_{n}}}\bigg)^{2}\alpha^{(2n)}_{\textbf{k}}+\exp{\left(-\frac{2ik}{k_{e_{n}}}\right)}\bigg(3i-\frac{2k^{3}}{k_{e_{n}}^{3}}+\frac{3ik^{2}}{k_{e_{n}}^{2}}\bigg)
\beta^{(2n)}_{\textbf{k}}\bigg],
\eea
\end{widetext} 
where horizon crossing conditions $-k_{s_{n}}c_{s}\tau_{s_{n}}=-k_{e_{n}}c_{s}\tau_{e_{n}}=1$ are employed to show explicit momentum dependence of the coefficients. 
It is important to note that the effective sound speed parameter $c_{s}$ in the current EFT context varies with conformal time. This information will be useful for future analysis discussed in this letter.
Consequently, the second Wilson coefficient $M_2$ appears in eqn.(\ref{action}) and also acquires a dynamical nature due to the parameter $c_{s}$. For the SR$_{1}$ phase, we fix $c_{s}(\tau_{*}) = c_{s}$ at the conformal time $\tau_{*}$ which corresponds to the pivot scale $k_{*}$. Let's examine the SR parameters for the multiple phases introduced earlier.

During the SR$_{1}$ phase, the first SR parameter $\epsilon$ gets treated as a constant until we reach the conformal time scale $\tau=\tau_{s_{1}}$. The second SR parameter $\eta = \epsilon^{\prime}/\epsilon{\cal H}$ is extremely small and behaves almost as a constant during SR$_{1}$.
As we encounter a sharp transition into the USR$_{1}$ phase, for the conformal time interval, $\tau_{s_{1}} \leq \tau \leq \tau_{e_{1}}$, the first SR parameter is described using the relation, $\epsilon(\tau) = \epsilon(\tau/\tau_{s_{1}})^{6}$, where $\epsilon$ is the SR parameter in the SR$_{1}$ phase. The second SR parameter displays a crucial behaviour in the USR$_{1}$ phase where it changes from $\eta\sim0$ to $\eta\sim-6$ at the conformal time, $\tau=\tau_{s_{1}}$, and changes again from $\eta\sim-6$ to $\eta\sim 0$ at the conformal time $\tau=\tau_{e_{1}}$ during exit from the USR$_{1}$ and entry into the next SR$_{2}$ phase. The SR$_{2}$ phase persists for the conformal time interval, $\tau_{e_{1}} \leq \tau \leq \tau_{s_{2}}$, where the first SR parameter satisfies the relation $\epsilon(\tau) = \epsilon(\tau_{e_{1}}/\tau_{s_{1}})^{6}$ displaying, in turn, a non-constant behaviour until the end of this phase where it undergoes another sharp transition into the next USR$_{2}$ phase at $\tau=\tau_{s_{2}}$, and the same pattern continues until inflation comes to an end.
The abrupt shifts in the value of $\eta$ make it essential to investigate enhancements in the scalar power spectrum amplitude. This crucial aspect of $\eta$ gets emphasized throughout the ongoing analysis done in this letter due to its severe impact on our various results.

In the present work, the choice of the wavenumber $k_{s_{1}}\sim{\cal O}(10^{4}\;{\rm Mpc}^{-1})$ is motivated by the need to incorporate the formation of large mass PBHs, $M_{\rm PBH}\sim{\cal O}(10^{4}\;M_{\odot})$, and to generate a GW spectrum consistent with the recently released NANOGrav 15 dataset which we discuss in the latter part of this letter. Based on the above choice, the wavenumber at the end of USR$_{1}$  is $k_{e_{1}}\sim{\cal O}(10^{5}\;{\rm Mpc}^{-1})$ to satisfy the e-folding constraint for this phase given by $\Delta{\cal N}_{\rm USR_{1}} \approx {\cal O}(2)$, as established by the authors in refs.\cite{Choudhury:2023vuj,Choudhury:2023jlt,Choudhury:2023rks,Choudhury:2023hvf}, which implies that a prolonged USR$_{1}$ is not allowed, to maintain the perturbative approximation during the large amplitude fluctuations. Throughout our analysis for the MSTs, we have consistently maintained a fixed interval for all USR phases to satisfy $k_{e_n}/k_{s_n}\sim{\cal O}(10\;{\rm Mpc^{-1}})$. This ensures that our theory maintains the perturbative approximations as $\Delta{\cal N}_{\rm USR_{n}} = \ln(k_{e_n}/k_{s_n}) \approx {\cal O}(2)\; \forall\;n$. To effectively manage the sharp nature of the transition during exit from the SR$_{2}$ phase and the entry into the next USR$_{2}$ phase, the total wavenumber interval of the SR$_{2}$ phase is carefully set to satisfy $k_{s_{2}}/k_{e_{1}}\sim{\cal O}(10^{3}\;{\rm Mpc}^{-1})$ which translates to e-foldings $\Delta{\cal N}_{\rm SR_{2}} \approx {\cal O}(7)$. With just having the SR$_{2}$ phase to end inflation, after taking into account the influence of one-loop contributions to the power spectrum associated with the scalar modes, it has been established by authors in ref.\cite{Choudhury:2023rks} that despite the implementation of RR procedures, the chosen transition wavenumbers across all three phases render successful inflation unattainable. To evade such constraints, we incorporate MSTs after the exit from the SR$_{2}$ and the entry into another USR$_{2}$ phase. Total e-foldings for the $n=1$ case turns out as $\Delta{\cal N}_{\rm SR_{1}} + \Delta{\cal N}_{\rm USR_{1}} + \Delta{\cal N}_{\rm SR_{2}} \approx {\cal O}(21)$, where $\Delta{\cal N}_{\rm SR_{1}} = \ln(k_{s_{1}}/k_{*}) \approx {\cal O}(12)$ from the mentioned values of $k_{*}$ and $k_{s_{1}}$ in the previous discussion. This calculation implies that a total of $n=6$ number of sharp transitions are required to satisfy the necessary condition for successful inflation of $\Delta{\cal N}_{\rm Total}\approx{\cal O}(60-70)$. Total e-foldings from the present MSTs construction give:
  \bea  
  \Delta{\cal N}_{\rm Total} = \sum^{7}_{n=1}\Delta{\cal N}_{\rm SR_{n}} +\sum^{6}_{n=1} \Delta{\cal N}_{\rm USR_{n}} \approx {\cal O}(66).
  \eea
The end of inflation occurs at the wavenumber $k_{\rm end}\sim{\cal O}(10^{27}\;{\rm Mpc}^{-1})$ corresponding to $\tau_{e_7}$, which also marks the end of the last SR$_{7}$ phase. 

Now, we are in a position to discuss the total tree-level scalar power spectrum. The information about the mode solutions in eqns.(\ref{solsr1},\ref{solusr},\ref{solsrn}) for the three different phases, SR$_{1}$, USR$_{n}$, and SR$_{n+1}$, is sufficient to give us the tree contribution to the power spectrum associated with the scalar modes in terms of the individual contributions as follows:
\begin{widetext}
    \bea \label{treeps}
\bigg[\Delta^{2}_{\zeta,{\bf Tree}}(k)\bigg]_{\bf Total} &=&\bigg[\Delta^{2}_{\zeta,{\bf Tree}}(k)\bigg]_{\textbf{SR}_{1}} \times \bigg[1+\sum^{6}_{n=1}\Theta(k-k_{s_{n}}) \left(\frac{k_{e_{n}}}{k_{s_{n}}}\right)^{6}\left|\alpha^{(2n)}_{\bf k}-\beta^{(2n)}_{\bf k}\right|^2\nonumber\\
&&\quad\quad\quad\quad\quad\quad\quad\quad\quad\quad\quad\quad\quad\quad\quad+\sum^{6}_{n=1}\Theta(k-k_{e_{n}})\times \left(\frac{k_{e_{n}}}{k_{s_{n}}}\right)^{6}\left|\alpha^{(2n+1)}_{\bf k}-\beta^{(2n+1)}_{\bf k}\right|^2\bigg].\quad\quad
\eea
\end{widetext}
where the tree-level SR$_{1}$ contribution is given by the expression:
\bea \bigg[\Delta^{2}_{\zeta,{\bf Tree}}(k)\bigg]_{\textbf{SR}_{1}} = \left(\frac{H^{2}}{8\pi^{2}M^{2}_{p}\epsilon c_s}\right)_{*} \left[1+\left(k/k_{s_{1}}\right)^2\right].
\eea
In eqn.(\ref{treeps}), second and third terms describe the USR$_{n}$ and SR$_{n+1}$ tree-level power spectrum, respectively, and $n=1,2,\cdots,6$ also label the relevant Bogoliubov coefficients for the $n$th phase as discussed before. The total tree-level scalar power spectrum involves a sum of the contributions from the regions where the sharp transitions are located and which we represent through the multiple Heaviside theta functions for each sharp transition. We can proceed further and introduce the one-loop corrections to the power spectrum associated with the scalar modes at the tree level coming from each region. To accomplish this requires using the perturbed action for the scalar modes computed at the third-order, which is given by:
\begin{widetext}
    \bea
    \label{s3action}
        S_{\zeta}^{(3)} &=& \int d\tau\;d^3x\;M_p ^2\;a^2\bigg [\bigg(3(c_s ^2 -1)\epsilon + \epsilon ^2 - \frac{\epsilon ^3}{2}\bigg )\; \zeta^{\prime} {^2} \zeta + \frac{\epsilon}{c_s ^2}\bigg(\epsilon - 2s +1 -c_s ^2 \bigg)(\partial_i \zeta)^2 \zeta  - \frac{2 \epsilon}{c_s ^2}\zeta^{\prime} (\partial_i \zeta) \bigg (\partial_i \partial ^{-2}\bigg(\frac{\epsilon \zeta^{\prime}}{c_s ^2}\bigg)\bigg) \nonumber \\ 
        && \quad \quad \quad \quad \quad \quad \quad \quad -\frac{\epsilon}{aH}\bigg (1-\frac{1}{c_s ^2}\bigg) \left(\zeta^{\prime} {^3}+\zeta^{\prime}(\partial_i \zeta)^2 \right) + \frac{\epsilon}{2}\zeta \bigg(\partial_i \partial_j \partial^{-2}\bigg (\frac{\epsilon \zeta^{\prime}}{c_s ^2}\bigg)\bigg)^2 + \underbrace{\frac{\epsilon}{2c_s ^2}\partial_{\tau} \bigg(\frac{\eta}{c_s ^2}\bigg)\zeta^{\prime} \zeta^2}_{\textbf{Dominant term}}+.....\bigg].
  \eea
\end{widetext}
where another SR parameter $s = c^{\prime}_{s}/{\cal H}c_{s}$ is introduced, $\eta = \epsilon^{\prime}/\epsilon {\cal H}$ is the usual second SR parameter, and $\cdots$ is used to denote the terms with highly suppressed contributions. The third-order action above can be obtained explicitly from the ADM formalism. For a thorough calculation of the action when $c_s=1$, especially in the context of single field inflation models, refer \cite{Maldacena:2002vr,Chen:2006xjb}. For computations with $c_{s} \ne 1$, refer \cite{Chen:2006nt,Chen:2010xka,Chen:2009zp}. The last term highlighted in the above action represents the dominant factor behind enhancements of loop corrections to the scalar power spectrum. This term contributes as ${\cal O}(\epsilon)$ for the USR$_{n}$ phases, while for the SR$_{1}$ and SR$_{n}$ phases it contributes as ${\cal O}(\epsilon^{3})$. The rest of the visible terms in the above action have contributions of ${\cal O}(\epsilon^{2})$ in the SR$_{1}$ and SR$_{n+1}$ and ${\cal O}(\epsilon^{3})$ in the USR$_{n}$ phases. The terms excluded after the dominant factor of the third order action are highly suppressed for each USR$_{n}$ phase and do not change the enhanced amplitude from the loop correction in any significant way. In the dominant term, a conformal time derivative is present which is another important reason for its significance during the encounter of the sharp transitions. The specifics of the nature of transition are also built into the parameters $\eta$ and $c_{s}$ with respect to conformal time.

Let us briefly discuss how the time-dependence is crucial for determining the behaviour of the corrections coming from the dominant term. The effective sound speed parameter $c_{s}(\tau)$ exhibits dynamical behaviour in the present EFT analysis. The effective sound speed parameter takes on a constant value $c_{s}$ during the SR$_{1}$ phase and for the conformal time interval during each USR and the rest of the SR phases present in the MSTs setup. This value remains almost constant and parameterized as:
\bea c_s(\tau)\sim c_s,\eea
but at the transition time scales it follows the parameterization:
\bea  c_{s}(\tau=\tau_{s_{n}}) \approx c_{s}(\tau=\tau_{e_{n}}) = \tilde{c}_{s}=1\pm\delta,\eea where $\delta$ being fine-tuned, such that $\delta \ll 1$ at the conformal time for the sharp transitions. This sudden non-trivial behaviour only at the transitions is encoded inside the dominant term $\partial_{\tau}(\eta/c^{2}_{s})$. Also in our MSTs construction, we observe subsequent enhancements of the scalar perturbations for different values of $n=1,2,\cdots,6$ due to the terms $\partial_{\tau}(\eta/c^{2}_{s}) \approx -\Delta\eta(\tau_{s_{n}})/\tilde{c}^{2}_{s}$ and $\partial_{\tau}(\eta/c^{2}_{s}) \approx \Delta\eta(\tau_{e_{n}})/\tilde{c}^{2}_{s}$ appearing at each conformal time scales $\tau=\tau_{s_{n}}$ and $\tau_{e_{n}}$ respectively. The approximation $\partial_{\tau}(\eta/c^{2}_{s})\approx 0$ is only valid during the SR$_{1}$, USR$_{n}$ and SR$_{n}$ period  but not at the sharp transition scales. From the above discussion, we describe the reason for the relevance of the dominant term in the USR$_{n}$ phase and why the remaining terms before this are only relevant for contributions in the SR$_{1}$ and SR$_{n+1}$ phases. To calculate the one-loop corrections to the power spectrum associated with the scalar modes, it becomes crucial to utilize $S_{\zeta}^{(3)}$ from eqn.(\ref{s3action}) for the necessary interaction Hamiltonian to evaluate the correlation functions. The interaction Hamiltonian can be constructed by the Legendre transformation of the third-order action. As such, we have ${\cal H} = - {\cal L}_ {int}^{(3)}$, where  ${\cal L}_{int}^{(3)}$ is the Lagrangian density obtained from the third order action. So, now we write the leading order Hamiltonian out of the action as :
\bea
\label{S3H}
H_{int}(\tau)&\supseteq &-\frac{M_p ^2}{2}\int d^3 x \; a^2 \bigg(\frac{\epsilon}{c_s ^2}\bigg)\partial_{\tau}\bigg(\frac{\eta}{c_s^2}\bigg)\zeta^{\prime}\zeta^2.
\eea
Since we require the last dominant term which has the highest contribution to the loop-corrected scalar power spectrum, we only write its corresponding self-interacting Hamiltonian. The remaining terms of the total Hamiltonian have relatively suppressed loop contributions in any USR$_{n}$ phase and therefore are not included in this discussion. Here we adopt the commonly-used Schwinger-Keldysh (in-in) formalism to calculate the loop corrections. 
Computing the desired one-loop 
corrections require evaluating the two-point function for the comoving curvature perturbation where Wick's theorem is implemented. 
We shall now present the final results for the one-loop contributions from the SR$_{1}$, USR$_{n}$, and SR$_{n+1}$ phases:
\begin{widetext}
    \bea \label{sr1correct}
\bigg[\Delta^{2}_{\zeta,\textbf{One-loop}}(k)\bigg]_{\textbf{SR}_{1}} &=& \bigg[\Delta^{2}_{\zeta,\textbf{Tree}}(k)\bigg]_{\textbf{SR}_{1}}^{2}
\times \left(1-\frac{2}{15\pi^{2}}\frac{1}{c_{s}^{2}k_{*}^{2}}\left(1-\frac{1}{c_{s}^{2}}\right)\epsilon\right)\times \left(c_{\textbf{SR}_{1}}-\frac{4}{3}{\cal J}_{\textbf{SR}_{1}}(\tau_{s_{1}})\right),
\\ \label{usrcorrect}
    \bigg[\Delta^{2}_{\zeta,\textbf{One-loop}}(k)\bigg]_{\textbf{USR}_{n}} &=&  \frac{1}{4}\bigg[\Delta^{2}_{\zeta,\textbf{Tree}}(k)\bigg]_{\textbf{SR}_{1}}^{2}
    \times \bigg\{ \bigg[ \bigg(\frac{\Delta\eta(\tau_{e_{n}})}{\tilde{c}^{4}_{s}}\bigg)^{2}{\cal J}_{\textbf{USR}_{n}}(\tau_{e_{n}}) - \left(\frac{\Delta\eta(\tau_{s_{n}})}{\tilde{c}^{4}_{s}}\right)^{2}{\cal J}_{\textbf{USR}_{n}}(\tau_{s_{n}})\bigg ]-c_{\textbf{USR}_{n}}\bigg\},\quad\quad
\\ \label{sr2correct}
\bigg[\Delta^{2}_{\zeta,\textbf{One-loop}}(k)\bigg]_{\textbf{SR}_{n+1}} &=& \bigg[\Delta^{2}_{\zeta,\textbf{Tree}}(k)\bigg]_{\textbf{SR}_{1}}^{2}
\times\left(1-\frac{2}{15\pi^{2}}\frac{1}{c_{s}^{2}k_{*}^{2}}\left(1-\frac{1}{c_{s}^{2}}\right)\epsilon\right)
\times \bigg(c_{\textbf{SR}_{n+1}}+{\cal J}_{\textbf{SR}_{n+1}}(\tau_{e_{n}})\bigg).
\eea
\end{widetext}

Here, the complete contribution from each phase requires evaluating the loop integrals represented by the terms ${\cal J}_{\textbf{SR}_{1}}(\tau_{s_{1}}),\; {\cal J}_{\textbf{USR}_{n}}(\tau_{s_{n}}),\;{\cal J}_{\textbf{USR}_{n}}(\tau_{e_{n}})$ and ${\cal J}_{\textbf{SR}_{n+1}}(\tau_{e_{n}})$ during the conformal time interval $\tau < \tau_{s_{1}}$, at the sharp transition time scale $\tau=\tau_{s_{n}}$ and $\tau = \tau_{e_{n}}$, and for the interval $\tau_{e_{n}} \leq \tau < \tau_{s_{n+1}}$ respectively. We also notice the use of the renormalization scheme dependent parameters labelled by $c_{\textbf{SR}_{1}},\;c_{\textbf{USR}_{n}}$ and $c_{\textbf{SR}_{n+1}}$ which are eventually fixed to some value during the renormalization procedure. We now mention the results for these integrals which are to be obtained under the late-time limit $\tau \rightarrow 0$:
\begin{widetext}
  \bea
{\cal J}_{\textbf{SR}_{1}}(\tau_{s_{1}}) &=& \lim_{\tau \to 0}\int_{k_{*}}^{k_{s_1}} \frac{dk}{k}g_{+,\textbf{k}}(\tau)g_{-,\textbf{k}}(\tau) = \lim_{\tau \to 0}\int_{k_{*}}^{k_{s_1}} \frac{dk}{k}\big|g_{+,\textbf{k}}(\tau)\big|^2= \lim_{\tau \to 0}\int_{k_{*}}^{k_{s_1}} \frac{dk}{k}\big|g_{-,\textbf{k}}(\tau)\big|^2\approx \ln\left(\frac{k_{s_1}}{k_{*}}\right),
\\
{\cal J}_{\textbf{USR}_{n}}(\tau_{s_{n}}) &=& \lim_{\tau\rightarrow 0}\int_{k_{s_{n}}}^{k_{e_{n}}}\frac{dk}{k}\bigg|\alpha^{(2n)}_{\bf k} g_{+,\textbf{k}}(\tau)-\beta^{(2n)}_{\bf k} g_{-,\textbf{k}}(\tau)\bigg|^2\approx\ln\left(\frac{k_{e_{n}}}{k_{s_{n}}}\right), 
\\
{\cal J}_{\textbf{USR}_{n}} (\tau_{e_{n}}) &=& \left(\frac{k_{e_{n}}}{k_{s_{n}}}\right)^{6}\lim_{\tau\rightarrow 0}\int_{k_{s_{n}}}^{k_{e_{n}}}\frac{dk}{k}\bigg|\alpha^{(2n)}_{\bf k}g_{+,\textbf{k}}(\tau)-\beta^{(2n)}_{\bf k}g_{-,\textbf{k}}(\tau)\bigg|^2\approx\left(\frac{k_{e_{n}}}{k_{s_{n}}}\right)^{6}{\cal J}_{\textbf{USR}_{n}}(\tau_{s_{n}}),\quad\quad
\\
{\cal J}_{\textbf{SR}_{n}}(\tau_{e_{n}}) &=& \left(\frac{k_{e_{n}}}{k_{s_{n}}}\right)^{6}\lim_{\tau\rightarrow 0}\int_{k_{e_{n}}}^{k_{s_{n+1}}}\frac{dk}{k}\bigg|\alpha^{(2n+1)}_{\bf k}g_{+,\textbf{k}}(\tau)-\beta^{(2n+1)}_{\bf k}g_{-,\textbf{k}}(\tau)\bigg|^2\approx \ln\left(\frac{k_{s_{n+1}}}{k_{e_{n}}}\right).
\eea
\end{widetext}

The results stated above were achieved through the implementation of the cut-off regularization procedure. The wavenumbers utilized in the upper and lower limits of the integrals signify the UV and IR cut-off scales, respectively. 
Another crucial fact to consider is the late-time limit $\tau \rightarrow 0$ taken for the loop contributions mentioned earlier. During the evaluation of the integrals associated with the one-loop Feynman diagrams, other terms are highly suppressed compared to the quadratic UV and logarithmic IR divergences. However, once we incorporate the late-time limit, the final results become free from quadratic UV divergences. So, we are left to deal with the logarithmic IR divergences from the loop integrals. The limiting method mentioned before yields identical results compared to the more rigorous Adiabatic/Wavefunction renormalization scheme, which is worked out in detail in ref.\cite{Choudhury:2023vuj}. Based on the above findings, only the logarithmic contributions remain which is an artifact of working with the QFT of de-Sitter space-time.

Before discussing a proper approach to handling the remaining logarithmic divergences, it is crucial to introduce the half-renormalized version of the one-loop corrected scalar power spectrum (OLCSPS):
\vspace{-0.5em}
\bea \label{unrenormps}
\Delta^{2}_{\zeta, {\textbf{EFT}}}(k) = \bigg[\Delta^{2}_{\zeta, {\textbf{Tree}}}(k)\bigg]_{\textbf{SR}_{1}}\bigg\{1+U+V\bigg\}.
\eea
Here $U$ and $V$ represent the total SR and USR contributions for MSTs. Using the eqns.(\ref{sr1correct}, \ref{usrcorrect}, \ref{sr2correct}) and the above-mentioned loop integrals we mention the following expressions:
\begin{widetext}
\bea
U&=& U_{\textbf{SR}_{1}} + U_{\textbf{SR}_{\textbf{rest}}},\\
U_{\textbf{SR}_{1}} &=&  -\frac{4}{3}\bigg[\Delta^{2}_{\zeta,\textbf{Tree}}(k)\bigg]_{\textbf{SR}_{1}}^{2}\times\left(1-\frac{2}{15\pi^{2}}\frac{1}{c_{s}^{2}k_{*}^{2}}\left(1-\frac{1}{c_{s}^{2}}\right)\epsilon\right)\ln\left(\frac{k_{s_1}}{k_{*}}\right),
\\
U_{\textbf{SR}_{\textbf{rest}}} &=& \sum^{6}_{n=1}U_{n} = \bigg[\Delta^{2}_{\zeta,\textbf{Tree}}(k)\bigg]_{\textbf{SR}_{1}}^{2}\times\sum^{6}_{n=1}\left(1-\frac{2}{15\pi^{2}}\frac{1}{c_{s}^{2}k_{*}^{2}}\left(1-\frac{1}{c_{s}^{2}}\right)\epsilon\right)\ln\left(\frac{k_{s_{n+1}}}{k_{e_{n}}}\right),
\\
V &=& \sum^{6}_{n=1}V_{n} = \frac{1}{4}\bigg[\Delta^{2}_{\zeta,\textbf{Tree}}(k)\bigg]_{\textbf{SR}_{1}}^{2}\times\sum^{6}_{n=1}\bigg(\frac{\left(\Delta\eta(\tau_{e_{n}})\right)^{2}}{\tilde{c}^{8}_{s}}\left(\frac{k_{e_{n}}}{k_{s_{n}}}\right)^{6} - \frac{\left(\Delta\eta(\tau_{s_{n}})\right)^{2}}{\tilde{c}^{8}_{s}}\bigg )\ln\left(\frac{k_{e_{n}}}{k_{s_{n}}}\right).
\eea
\end{widetext}
The expression in eqn.(\ref{unrenormps}) includes logarithmic contributions, which need to be addressed to obtain a physically relevant result for the OLCSPS. To tackle this, our strategy revolves around renormalizing the obtained power spectrum to soften the remaining logarithmic divergences. 
The softening of the IR divergences invites the use of a counter-term for the half-renormalized scalar power spectrum. This approach towards renormalization gets performed $6$ times, accounting for the loop contributions due to the SR and USR phases, which accommodate the MSTs in the present setup.

In light of this discussion, we introduce the power spectrum renormalization scheme. The pursuit of this approach leads us to define the following re-scaled quantity,   
the tree level power spectrum of SR$_{1}$ as $\small[\Delta_{\zeta,\textbf{Tree}}^{2}(k)\small]_{\textbf{SR}_{1}} = 6 \times \small[\widetilde  \Delta_{\zeta,\textbf{Tree}}^{2}(k)\small]_{\textbf{SR}_{1}}$.
The necessity for this re-scaling will become clear shortly. In terms of these new quantities, we begin by writing the half-renormalized scalar power spectrum which includes a sharp transition:
\bea
\label{nthUnPS}
\big[\Delta^{2}_{\zeta, {\textbf{EFT}}}(k)\big]_{n} = \big[\widetilde \Delta^{2}_{\zeta, {\textbf{Tree}}}(k)\big]_{\textbf{SR}_{1}}\big\{1 +  W_{n} + 6V_{n} \big\},\quad
\eea
where $W_{n} = U_{\rm \textbf{SR}_{1}} +  6U_{n}$. Here $n$ acts as a counter for the number of sharp transitions while $6$ represents the total number of MSTs in the present context. Using the above definition, we write the total half-renormalized scalar power spectrum as:
\bea
\Delta^{2}_{\zeta, {\textbf{EFT}}}(k) &=&  \sum^{6}_{n=1}\big[\Delta^{2}_{\zeta, {\textbf{EFT}}}(k)\big]_{n}
\eea
In terms of the rescaled quantity, summing eqn.(\ref{nthUnPS}) over terms for the $6$ MSTs, we get the total half-renormalized power spectrum:
\vspace{-1em}
\bea
\Delta^{2}_{\zeta, {\textbf{EFT}}}(k) =  \big[\widetilde \Delta^{2}_{\zeta, {\textbf{Tree}}}(k)\big]_{\textbf{SR}_{1}}\sum^{6}_{n=1}\big[1+ W_n +6V_n \big].
\eea
Now we begin by writing the renormalized version for the $n$th contribution of the scalar power spectrum by introducing the necessary counter-term as, $\overline{\big[\Delta_{\zeta, \textbf{EFT}}^{2}(k)\big]}_{n} = {\cal Z}_{n}^{\rm IR} \; \big[\Delta_{\zeta,\textbf{EFT}}^{2}(k)\big]_{n}$, where ${\cal Z}_{n}^{\rm IR}$ is the counter-term also known as the renormalization factor. This factor is determined explicitly from the renormalization condition using the pivot scale value $k_{*}$ as, $\big[\overline{\Delta_{\zeta, \textbf{EFT}}^2 (k_{*})}\big]_{n} = \big[\widetilde \Delta_{\zeta, \textbf{Tree}}^2 (k_{*})\big]_{\textbf{SR}_{1}}$. The condition stated before results in the following form of the counter-term:
\bea
{\cal Z}_{n}^{\rm IR}= \frac{\big[\widetilde\Delta_{\zeta, \textbf{Tree}}^2 (k_{*})\big]_{\textbf{SR}_{1}}}{\big[\Delta_{\zeta, \textbf{EFT}}^2 (k_{*})\big]_{n}}\approx (1- W_{n,*}-6V_{n,*}+\cdots).\quad
\eea
Upon using this determined counter-term we can write the total renormalized OLCSPS for the case of $6$ MSTs:
\bea
\overline{\Delta^{2}_{\zeta, {\textbf{EFT}}}(k)} = \big[ \Delta^{2}_{\zeta, {\textbf{Tree}}}(k)\big]_{\textbf{SR$_{1}$}}\times\left\{1+\frac{1}{6}\sum^{6}_{n=1}{\cal Q}_{n,\textbf{EFT}}\right\},\quad
\eea
where we define ${\cal Q}_{n,\textbf{EFT}}$ as:
\bea
{\cal Q}_{n,\textbf{EFT}} =  \frac{-6\big[\Delta_{\zeta, \textbf{Tree}}^2 (k)\big]_{\textbf{SR}_{1}}}{\big[\Delta_{\zeta, \textbf{Tree}}^2 (k_{*})\big]_{\textbf{SR}_{1}}}\left\{W^{2}_{n,*} + 36V^{2}_{n,*} + \cdots \right\}.\hspace{0.32cm}
\eea
The frequent use of $6$ in our calculations represents the necessary number of MSTs in the present context to satisfy successful inflation. Any other number greater than $6$ for the total number of MSTs can also appear in the above results. For instance, if our case required $10$ MSTs, then the factor of $6$ is replaced by $10$ throughout, and the conclusions will remain invariant.
Now, we invoke the resummation procedure over the obtained results for the renormalized OLCSPS to finally obtain a finite output, which also ends up validating the cosmological perturbative approximations. The quantity ${\cal Q}_{n,\textbf{EFT}}$ has to be obtained in a manner to satisfy the said approximation. The resummation is done with the help of the Dynamical Renormalization Group (DRG) scheme \cite{Boyanovsky:1998aa,Boyanovsky:2001ty,Boyanovsky:2003ui,Burgess:2009bs,Dias:2012qy,Chen:2016nrs,Baumann:2019ghk,Burgess:2009bs,Chaykov:2022zro,Chaykov:2022pwd,Choudhury:2023vuj,Choudhury:2023jlt,Choudhury:2023rks} which is an improved version of the RG resummation method pertaining to a specific range of momentum scales that are observationally attainable and lie within the small coupling regime. In this approach, the scale-dependent cosmological beta functions absorb the secular time-dependent contributions of the theory, thereby validating the perturbative expansion. This technique is only viable if the corresponding resummed series converges at a late time limiting conformal time scales $\tau \to 0$. 
    \begin{figure*}[htb!]
    	\centering
   {
   \includegraphics[width=18cm,height=7.5cm] {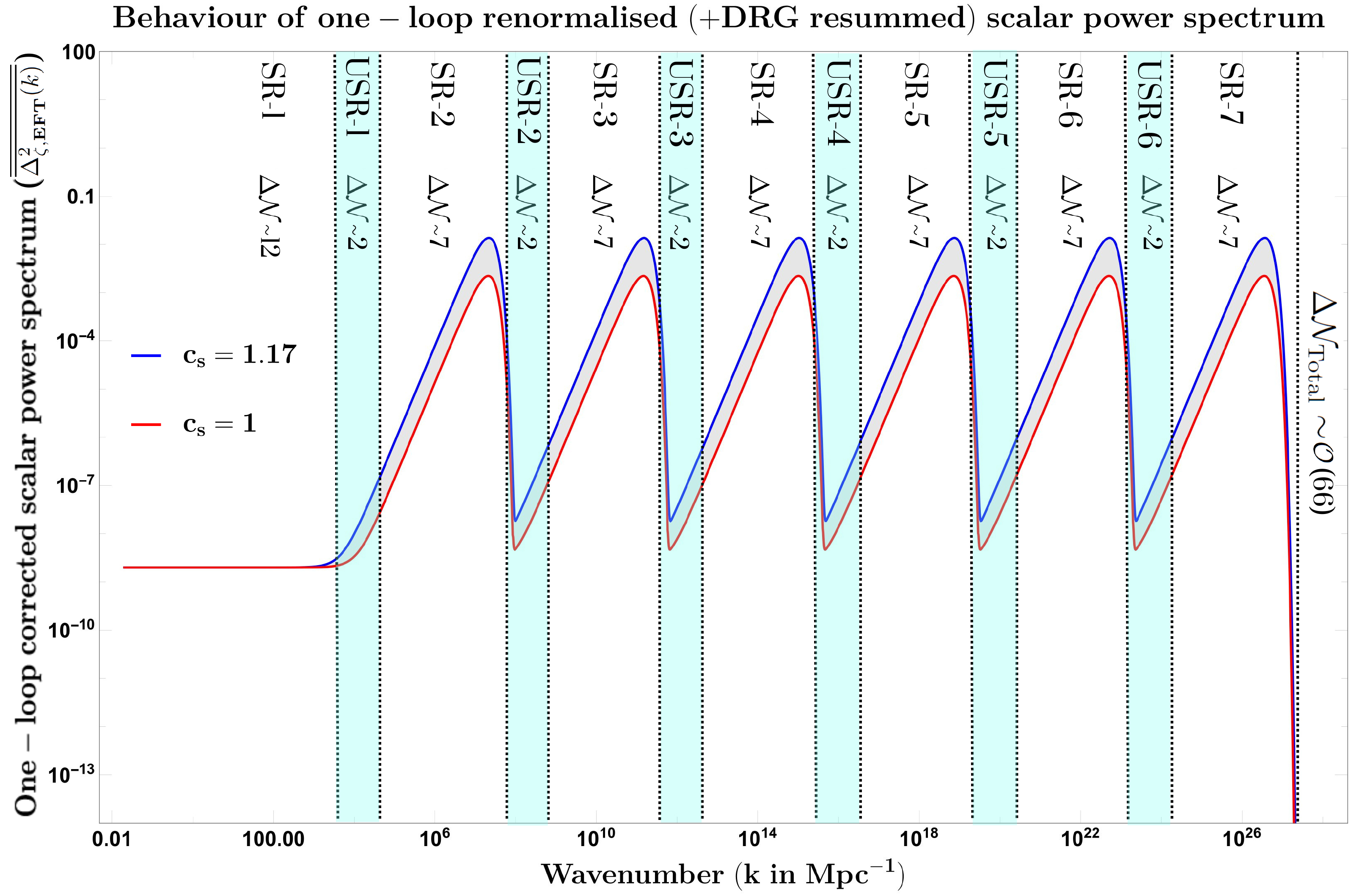}
    }
    	\caption[Optional caption for list of figures]{The above plot displays the one-loop corrected scalar power spectrum, which is renormalized and resummed through the DRG technique, across a wide range of wavenumbers. The plot features clear markers (cyan bands) for each ultra-slow roll (USR$_{n}$) and slow-roll (SR$_{n+1}$) phase, labelled for all $n=1,2,\cdots,6$. The behaviour of the spectra is shown for two cases of effective sound speed: $c_{s}=1$ (red) and $c_{s}=1.17$ (blue). The grey band between the two spectra represents the acceptable parameter range for the amplitude of enhancements, which should be between ${\cal O}(10^{-3}-10^{-2})$ for PBH production.} 
    	\label{final1}
    \end{figure*}
Performing the DRG resummation method yields the following final result in the present context of MSTs\footnote{By replacing the number $6$ by $1$, one can confirm that the modified results boil down to the one-loop renormalized and resummed scalar power spectrum (OLRRSPS) for a single sharp transition as obtained in refs \cite{Choudhury:2023vuj,Choudhury:2023jlt,Choudhury:2023rks}.}:
\begin{widetext}
 \bea \label{finalps}
 \overline{\overline{\Delta^{2}_{\zeta, {\textbf{EFT}}}(k)}}
 = \big[\Delta^{2}_{\zeta, {\textbf{Tree}}}(k)\big]_{\textbf{SR}_{1}}\times\bigg\{\exp{\left({\cal Q}_{1,\textbf{EFT}}\right)}+\sum^{6}_{n=2} \Theta(k-k_{s_{n}})\exp{\left({\cal Q}_{n,\textbf{EFT}}\right)}\bigg\}\times \bigg(1+{\cal O}\big[\Delta_{\zeta, \textbf{Tree}}^2 (k_{*})\big]_{\textbf{SR}_{1}}\bigg).\quad
 \eea
\end{widetext}

Fig.(\ref{final1}) represents the behaviour of the one-loop renormalized and DRG resummed scalar power spectrum for a wide range of wavenumbers, which are important for our analysis in the latter half of this letter. The figure depicts the case of MSTs analyzed for two different values of the effective sound speed, namely $c_s=1$ (red) and $c_{s}=1.17$ (blue). The region between the two spectra represents the allowed parameter space of our theory. It is minute but crucial as its significance lies in the fact that not only does the perturbation theory hold good within it, but outcomes derived from this region also contribute to the generation of SIGWs and yield sufficient abundance range for the formation of dark matter.
The value of the pivot scale is set to $k_{*} = 0.02{\rm Mpc}^{-1}$, and intervals for the subsequent USR and SR phases are managed to satisfy $\Delta{\cal N}_{\rm USR_{n}}\approx{\cal O}(2)$ and $\Delta{\cal N}_{\rm SR_{n+1}}\approx{\cal O}(7)$ for $n=1,2,\cdots,6$, respectively. The plot shows that mildly breaking the causality constraint by choosing $c_{s}=1.17$ is a preferred scenario since it generates the sufficient amplitude of ${\cal O}(10^{-2})$ desired for the production of PBH. The value of $c_{s}=1.17$ is also unique for being the value above which the perturbative approximations completely break down, and this fact is discussed in detail by the authors in refs.\cite{Choudhury:2023jlt,Choudhury:2023rks}. Also, carefully examining the case of $c_{s} < 1$, for instance, $c_{s}=0.6$, reveals that the amplitude of the power spectrum remains significantly lower than the threshold required for PBH formation, even though the perturbative approximations hold good. The case with $c_{s} < 1$ is discussed in detail in refs.\cite{Choudhury:2023jlt,Choudhury:2023rks}. Here we must emphasize that a slight increase in the value of the parameter from $c_{s}=0.6$ to $1$, leads to a substantial enhancement in the amplitude of the OLRRSPS. The figure displays consecutive peaks and dips which show consistency in their respective amplitudes throughout the spectrum. The peak values of the amplitude lies within, ${\cal O}(10^{-3}-10^{-2})$, and the dip values of the amplitude lies, ${\cal O}(10^{-8}-10^{-9})$ for effective sound speed, $1 \leq c_{s} \leq 1.17$. The final peak amplitude after the last transition into SR$_{7}$ phase falls sharply at ${\cal O}(10^{27}{\rm Mpc}^{-1})$ as a result of the strong DRG resummation procedure. 
This kind of power spectra generated resemble multiple bumpy-like features in the potential as proposed in \cite{Mishra:2019pzq}. We have carried out a model-independent study under the EFT framework without considering any specific structure in the effective potential. In terms of the MSTs, we have included the features in the second SR parameter $\eta$ which can suffice the same purpose in the present context. However, in our future follow-up projects, we will use a specific class of effective potentials in the framework of EFT with the canonical and non-canonical single-field inflation models in detail.

The OLRRSPS contains informative details about the fraction of energy density of PBHs as compared to dark matter. It will also be used to generate SIGWs that will cover a wide range of frequencies. All these will be discussed in the latter half of the letter. In between the discussions, let us also remind ourselves that including multiple phases of SR and USR comes with its challenges. Recall that we are dealing with large quantum fluctuations from $6$ USR phases, and each such phase is followed up by another SR phase, which demands us to adjust the amplitude of the power spectrum for the scalar modes according to the expected behaviour in each phase. 
One also has to consider the validity of the cosmological perturbation throughout the scales involved in the theory covering the CMB pivot scale of $k_{*}\sim{\cal O}(0.02\;{\rm Mpc}^{-1})$ to the end of inflation scale $k_{\rm end}\sim{\cal O}(10^{27}\;{\rm Mpc}^{-1})$. 

This letter also addresses constrained outcomes from previous endeavours. In the earlier works \cite{Choudhury:2023vuj,Choudhury:2023jlt,Choudhury:2023rks}, the authors have explicitly shown that a strong No-go theorem exists for the PBH mass according to which large mass PBHs cannot be generated using the OLRRSPS to achieve the sufficient e-foldings of $\Delta{\cal N}_{\rm Total}\approx{\cal O}({60-70})$ in the presence of single sharp transition. In the ref.\cite{Choudhury:2023rks}, the authors have pointed out that if the PBH mass is $M_{\rm PBH}\sim{\cal O}(M_{\odot})$, then the total e-foldings is $\Delta{\cal N}_{\rm Total}\approx{\cal O}(25)$, which is not at all sufficient to achieve successful inflation with the corresponding theoretical set-up. On the other hand, to maintain such a strong requirement on $\Delta{\cal N}_{\rm Total}$ to end inflation, only $M_{\rm PBH}\sim 10^{2}\;{\rm gm}$ can be obtained from this setup with a single sharp transition at the high momenta scales, which is not very interesting from a cosmological perspective. Therefore, we have incorporated $6$ consecutive MSTs designed in the SR, USR, and SR sequence. In this setup, we have explicitly proven that the strong No-go in the obtained PBH mass can be evaded without losing any generality. As a consequence of the performed analysis presented in this letter, we can cover the whole range of wave numbers $k\sim{\cal O}(10^{-2}\;\rm Mpc^{-1} - 10^{27}\;{\rm Mpc^{-1}})$ while also achieving the sufficient e-foldings required to end inflation. 
This implies that by accommodating $6$ peaks of large amplitude fluctuations which lie in the order of ${\cal O}({10^{-3}-10^{-2}})$ for $1 \leq c_{s} \leq 1.17$, one can cover the whole spectrum of PBH mass, $M_{\rm PBH}\sim{\cal O}(10^{-31}M_{\odot}-10^{4}M_{\odot})$. So, such an engineered set-up not only evades the strong constraints on the PBH mass as coming from renormalization and DRG resummation but also explains the generation of SIGWs in the frequencies ranging from $f\sim{\cal O} (10^{-17}{\rm Hz}-10^{12}{\rm Hz})$. 
With the present methodology from the EFT of single field inflation itself, one can not only explain the enhancement of the GW spectra as observed in the NANOGrav $15$ data but also respects the constraints from LISA, LIGO, ET, CE, BBO, HLVK, and various other terrestrial and space-based gravitational experiments. Upon reflection, we arrive at the recognition that this discovery is noteworthy in light of the ongoing debate that has persisted for several months. Nevertheless, this is not the sole accomplishment in which the formidable No-go theorem can be evaded as proposed in ref \cite{Choudhury:2023vuj,Choudhury:2023jlt,Choudhury:2023rks}. Recently some authors of this paper have explicitly shown that the No-go theorem of the PBH mass can be evaded with the help of the Non-renormalization theorem under the setup of USR Galileon inflation, where the generalized Galilean shift symmetry is softly broken. See refs. \cite{Choudhury:2023hvf,Choudhury:2023kdb,Choudhury:2023hfm} for more details on this issue. So, it not only helps to generate PBH masses of all ranges, but can also produce the desired SIGWs which satisfies the constraints of NANOGrav $15$, and other terrestrial and space-based gravitational experiments.

Furthermore, all of these frameworks mentioned have been designed with the help of sharp transitions \cite{Choudhury:2023hvf,Choudhury:2023kdb,Choudhury:2023hfm,Kristiano:2022maq,Riotto:2023hoz,Kristiano:2023scm,Franciolini:2023lgy,Cheng:2023ikq,Motohashi:2023syh,Riotto:2023gpm, Firouzjahi:2023ahg,Firouzjahi:2023aum,Franciolini:2023lgy,Cheng:2023ikq,Tasinato:2023ukp,Motohashi:2023syh}. 
However, one can also integrate smooth transitions \cite{Riotto:2023gpm,Firouzjahi:2023ahg,Firouzjahi:2023aum}. Notably, the authors in \cite{Riotto:2023gpm,Firouzjahi:2023ahg,Firouzjahi:2023aum} have pointed out that employing a single smooth transition can offer a solution to evade the challenges due to large quantum fluctuations and generate huge PBH masses while explaining the production of SIGWs. The impact of smooth transitions remains debated in the literature due to the need for comprehensive RR to support the theory. Nonetheless, it would be crucial to seek such possibilities for future work. On the contrary, our letter is a comprehensive investigation into the utility of MSTs using techniques such as renormalization and DRG resummation. We have successfully overcome all the drawbacks mentioned above through meticulous computation in the present work.

The following discussion based on the fractional abundance of PBHs provides an estimated idea of the PBH masses obtained corresponding to specific values for the threshold of density contrast and the transition wavenumbers in the OLRRSPS.
Within the EFT framework, it is possible to provide the above estimate with a dependence on the parameter $c_{s}$ by employing the following expression:
\bea
\label{fmass}
\bigg[ \frac{M_{\rm {PBH}}}{M_{\odot}}\bigg]&=&1.13\times 10^{15}\nonumber\\
&\times& \bigg[\frac{\gamma}{0.2}\bigg]\bigg[\frac{g_{*}}{106.75}\bigg]^{-\frac{1}{6}} \left(\frac{k_{*}}{k_{\rm PBH}}\right)^{2}c_{s}^{2}.
\eea
where the efficiency factor $\gamma\sim 0.2$, and the total relativistic degrees of freedom in the SM is $g_{*} = 106.75$. The modification due to the underlying EFT setup is seen clearly from the use of the parameter $c_{s}.$ The fractional abundance of PBHs provides an estimated mass range allowed for the PBHs in the current MSTs setup. This requires using the OLRRSPS and the PBH mass fraction at formation as: 
\bea
\label{fPbh}
f_{\rm PBH}&=&1.68\times 10^8 \times \bigg[\frac{\gamma}{0.2}\bigg]^{\frac{1}{2}}\times \bigg[\frac{g_{*}}{106.75}\bigg]^{-\frac{1}{4}}\nonumber\\
&\times& \bigg[\frac{M_{\rm PBH}}{M_{\odot}}\bigg]^{-\frac{1}{2}}\times \beta(M_{\rm PBH}).
\eea
where the mass fraction $\beta(M_{\rm PBH})$ corresponds to the probability of the event where the coarse-grained density contrast exceeds a certain threshold on its value $\delta_{\rm th} > \delta_{\rm c} = 1/3.$ The required probability is calculated from the Press-Schechter formalism as:
\bea
\label{beta}
\beta \approx \gamma \bigg[\frac{\sigma_{M_{\rm PBH}}}{\sqrt{2 \pi}\delta_{\rm th}}\bigg]\rm exp \bigg(-\frac{\delta_{th}^2}{2\sigma_{M_{\rm PBH}}^2}\bigg).
\eea
Here, the quantity $\delta_{\rm th}$ is the coarse-grained density contrast after using a Gaussian window function over the scales of PBH formation $R=1/(\tilde{c_{s}}k_{\rm PBH})$. The corresponding variance needed to perform this calculation is given as:
\bea
\label{sigma}
\sigma_{M_{\rm PBH}}^2
= \frac{16}{81}\int_{0}^{\infty}\frac{dk}{k}\;(kR)^{4} e^{-k^{2}R^{2}/2}\;\overline{\overline{\Delta^{2}_{\zeta, {\textbf{EFT}}}(k)}}.
\eea
This further requires the well-known relation for the power spectrum of the density contrast which involves the previously obtained result in eqn.(\ref{finalps}).
Here, we must emphasize to eliminate any confusion that the above relation for the density contrast works for the linear regime and does not incorporate any of the existing non-linearities coming from the spatial derivatives in the comoving curvature perturbation.
    \begin{figure}[htb!]
    	\centering
 \subfigure[For $c_{s}=1$ and $c_{s}=1.17$ with the chosen values of the threshold density contrast as $\delta_{\rm th}=0.4,0.66.$]{
      	\includegraphics[width=7.8cm,height=5.5cm]{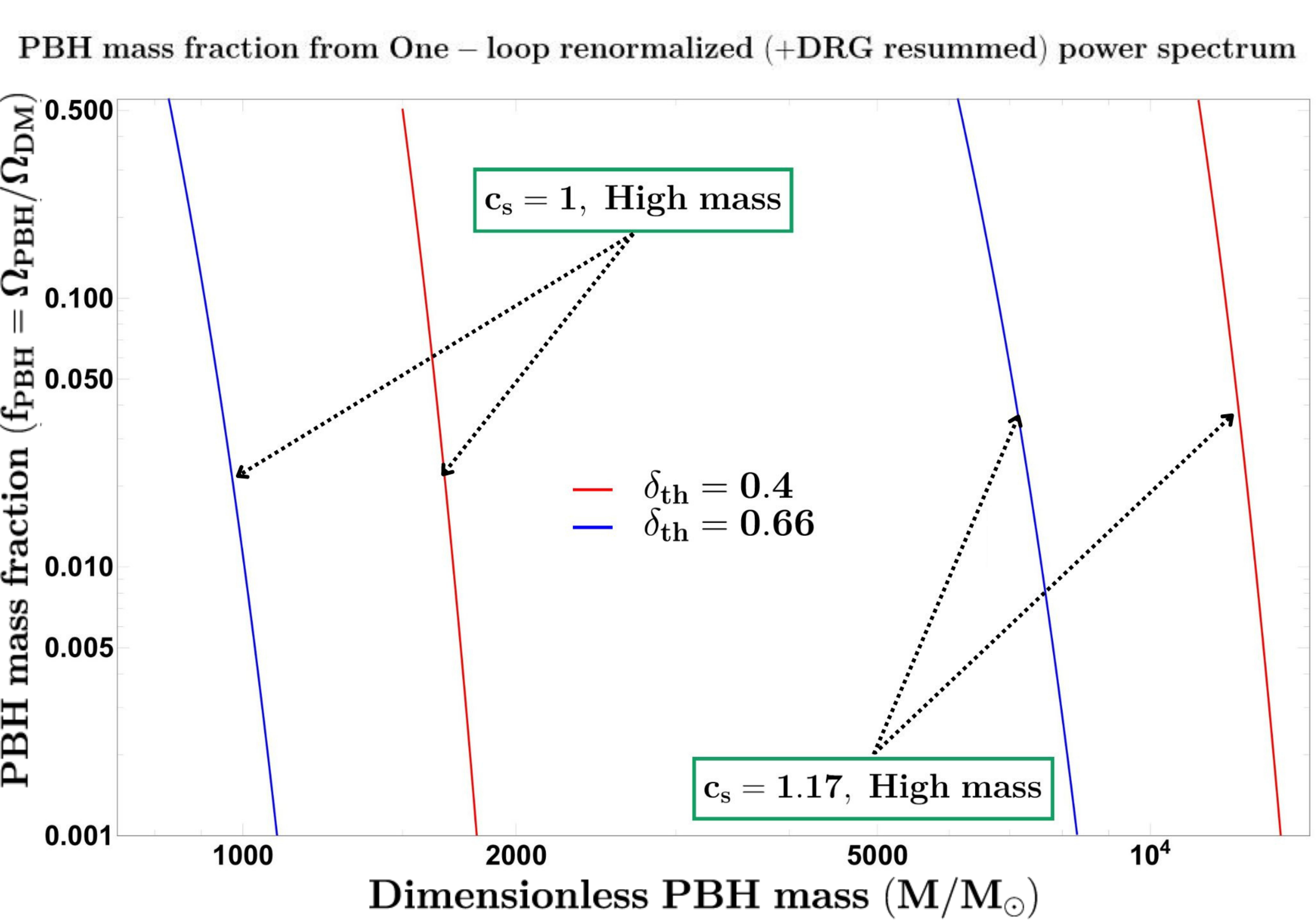}\label{N1}}
    \subfigure[For $c_{s}=1$ and $c_{s}=1.17$ with the chosen values of the threshold density contrast as $\delta_{\rm th}=0.4,0.47,0.48.$]{
       \includegraphics[width=7.8cm,height=5.5cm]{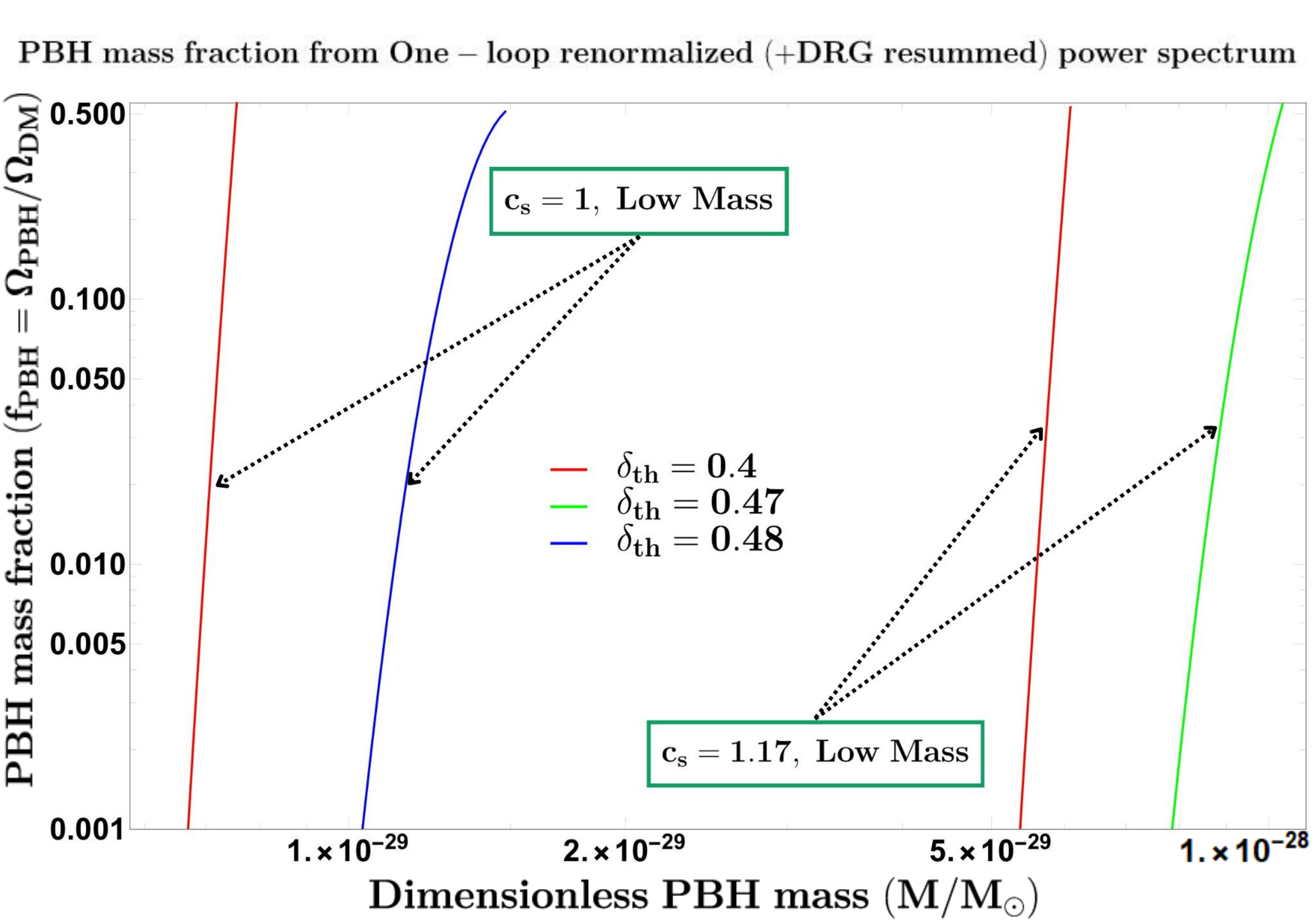}\label{N2}}
       
    	\caption[Optional caption for list of figures]{The plot above illustrates the fractional abundance of PBHs, separated into high and low masses. Each plot shows the effective sound speed parameter, which satisfies $c_{s}=1$ and $c_{s}=1.17$.} 
    	\label{final2}
     \end{figure}
Fig.(\ref{N1}-\ref{N2}) depicts the allowed range of PBH masses which lie within the interval, $0.001 \lesssim f_{\rm PBH} \lesssim 0.5$, and provides a fractional abundance of PBHs, which also signifies the observationally relevant dark matter density as measured today. We have provided the plots for two separate cases of the effective sound speed, $c_{s}=1$ and $c_{s}=1.17$, and analyzed the abundance behaviour by keeping the threshold density contrast within the interval $\delta_{\rm th} \in [2/5, 2/3]$ \cite{Musco:2018rwt}. The threshold interval is chosen to maintain the perturbative approximation within the computation of the present quantity. In \ref{N1}, for the case of $\delta_{\rm th}=0.4,0.66$, we found that the allowed mass range of heavy PBHs satisfies $M_{\rm PBH}\sim{\cal O}(10^{3}M_{\odot}-10^{4}M_{\odot})$ which corresponds to the masses obtained for the first sharp transition scale at $k_{\rm PBH}\sim{\cal O}(10^{4}{\rm Mpc}^{-1})$. In \ref{N2}, we analyze the abundance behaviour for the specific values of the threshold density contrast $\delta_{\rm th}=0.4,0.47,0.48$. After adopting the said values of the threshold, we obtain the required abundance for the small mass PBHs which satisfies, $M_{\rm PBH}\sim{\cal O}(10^{-29}M_{\odot}-10^{-28}M_{\odot})$, and correspond to the transition scale $k_{\rm PBH}\sim{\cal O}(10^{19}{\rm Mpc}^{-1})$. For the case where the threshold of the density contrast is pushed to higher values until we reach its upper bound from the interval, we encounter negligibly small values for the fractional abundance of PBHs, even for cases where the PBH masses are significantly lower relative to our analysis for the low mass PBHs case.
We expect that due to having multiple USR phases in our present MST setup, one can address the problems of PBH overproduction \cite{Choudhury:2023fwk,Choudhury:2023hfm,Franciolini:2023pbf,Inomata:2023zup,Wang:2023ost,Balaji:2023ehk,HosseiniMansoori:2023mqh,Gorji:2023sil} by calculating the three-point functions and the bispectrum from the scalar comoving curvature perturbation as computed in \cite{Choudhury:2023kdb}. In this letter, we have not included the non-linearities and associated non-Gaussianities \cite{Musco:2020jjb,DeLuca:2023tun,Harada:2015yda,DeLuca:2019qsy,Young:2019yug}. However, we plan to address these aspects in our future projects shortly. The outcomes obtained from the Press-Schechter theory will not change drastically if we demand that the cosmological perturbative approximation holds good throughout the evolution in these consecutive USR phases, which is the necessary key ingredient to produce PBH masses with the corresponding momenta scales. 

    \begin{figure*}[htb!]
    	\centering
   {
      	\includegraphics[width=19cm,height=7.5cm] {
      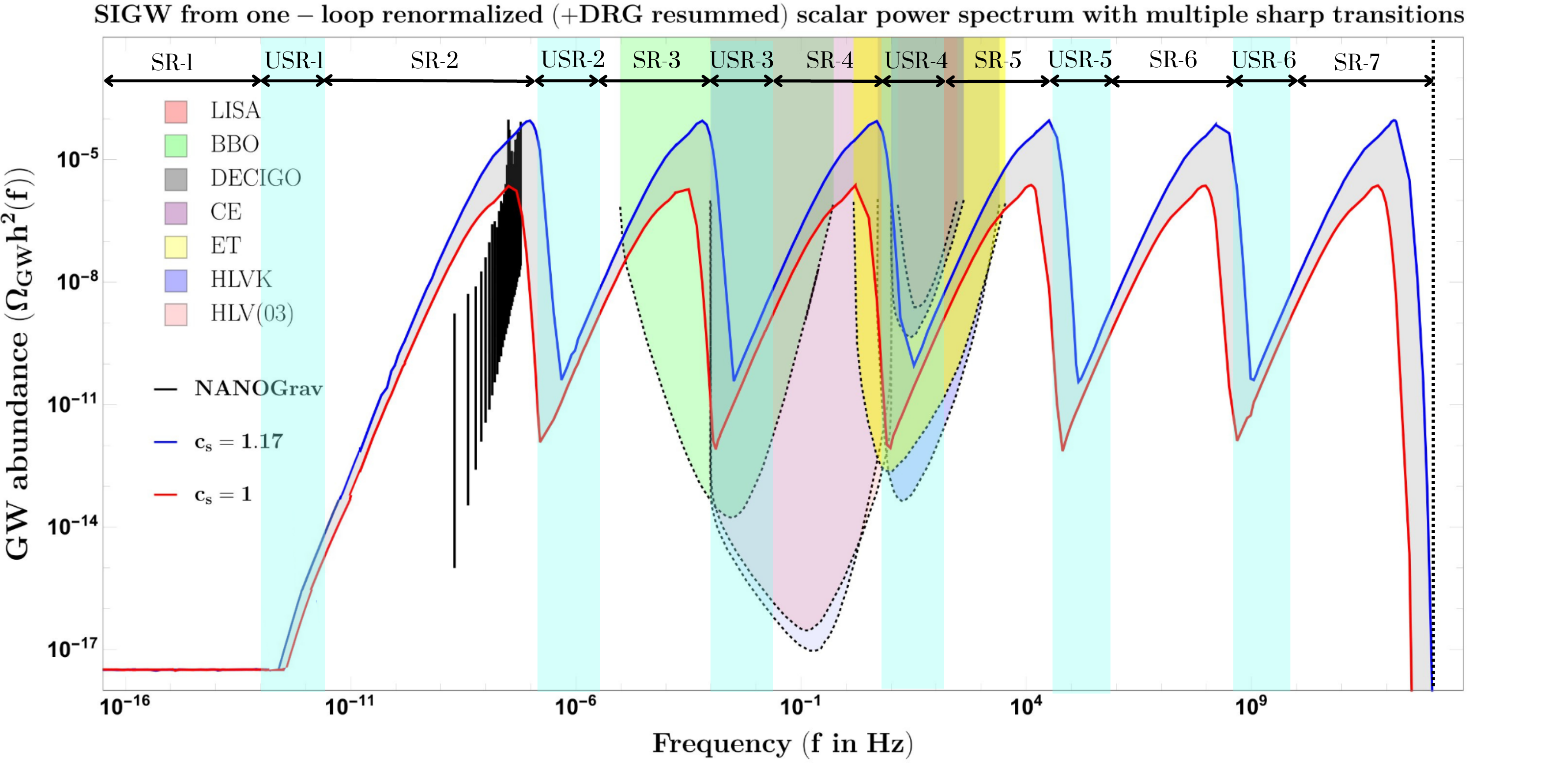}
        \label{K3}
    }
    	\caption[Optional caption for list of figures]{The plot depicts the GW spectra generated from the one-loop corrected scalar power spectrum, which is renormalized and resummed through the DRG technique, with $c_s=1$ (red) and $c_s=1.17$ (blue). The plot features clear markers (cyan bands) for each ultra-slow roll (USR$_{n}$) and slow-roll (SR$_{n+1}$) phase, labelled for all $n=1,2,\cdots,6$. The vertical lines spanning the frequency range of $f\sim{\cal O}(10^{-9}{\rm Hz}-10^{-7}{\rm Hz})$ indicate the NANOGrav $15$ data. Notably, the first peak of the GW spectra aligns with the peak of the NANOGrav $15$ data, and the remaining signal is within the projected sensitivity ranges of both current and forthcoming experiments designed to identify high-frequency gravitational waves. These experiments encompass LISA, DECIGO, BBO, Einstein Telescope (ET), Cosmic Explorer (CE), the HLVK network (comprising detectors such as aLIGO in Hanford and Livingston, aVirgo, and KAGRA), and HLV (O3).}

    	\label{finalGW}
    \end{figure*}

Efforts have been going on to fit the SIGW spectra and other source-generated GW with NANOGrav $15$ data for quite a while. See refs. \cite{Choudhury:2023hfm,Franciolini:2023pbf,Inomata:2023zup,Wang:2023ost,Balaji:2023ehk,HosseiniMansoori:2023mqh,Gorji:2023sil,DeLuca:2023tun,Choudhury:2023kam,Yi:2023mbm,Cai:2023dls,Vagnozzi:2023lwo,Frosina:2023nxu,Zhu:2023faa,Jiang:2023gfe,Cheung:2023ihl,Oikonomou:2023qfz,Liu:2023ymk,Wang:2023len,Zu:2023olm, Abe:2023yrw, Gouttenoire:2023bqy, Xue:2021gyq, Nakai:2020oit, Athron:2023mer, Madge:2023cak,Kitajima:2023cek, Babichev:2023pbf, Zhang:2023nrs, Zeng:2023jut, Ferreira:2022zzo, An:2023idh, Li:2023tdx,Ellis:2020ena, Blasi:2020mfx, Vagnozzi:2020gtf, Chen:2019xse, Madge:2023cak, Liu:2023pau, Yi:2023npi, Liu:2023ymk, Buchmuller:2021mbb} for more details. Apart from having various sources of SIGW, we have directed our focus on the SIGW spectra and enhancement of the GW signal at the MSTs points which can not only converge well with the NANOGrav $15$ data but also satisfy the constraints of the other probes such as the LISA, LIGO, and other observatories. 
We aim to achieve this without altering any theoretical parameters. Instead, we employ a single parameter and leverage the MSTs to achieve our target. This task presents stringent challenges due to the strong No-go constraints as previously discussed in our letter. Not only have we evaded the No-go theorem but also obtained distinct GW spectra that span a broad frequency range, extending from $f\sim{\cal O}(10^{-17}{\rm Hz} - 10^{12}{\rm Hz})$. It is noteworthy to mention that the peak amplitudes of the OLRRSPS are kept consistent without compromising the validity of the perturbation theory in primordial cosmological scales.

In this framework, the scalar counterpart in the spatially flat FLRW background metric acts as a source for producing the second-order perturbations of the tensor modes. The SIGWs are manifestations of these tensor perturbations. Mentioned below are the fundamental equations that are responsible for the generation of the SIGWs:

\begin{widetext}
\bea
\label{Omegagw}
\Omega_{\rm GW}(\tau ,k) &=& \frac{\rho_{\rm GW}(\tau,k)}{\rho_{\rm tot}(\tau)}=\frac{1}{24}\times\bigg[\frac{k}{a(\tau)H(\tau)}\bigg]^2 \; \times\langle \overline{\overline{\Delta_h ^2 (\tau,k)}}\rangle,
\\
\label{Omegaf}
\Omega_{\rm GW}(f) &=& 0.39\times\left[\frac{g_{*}(T_{\rm rad})}{106.75}\right]^{-1/3}\Omega_{r,0}\times \int^{\infty}_{0}dy\int^{1+y}_{|1-y|}dx\;{\cal G}(x,y)\times\overline{\overline{\Delta_{\zeta,\textbf{EFT}}^{2}(kx)}}\times \overline{\overline{\Delta_{\zeta,\textbf{EFT}}^{2}(ky)}},
\\
\label{kernel}
{\cal G}(x,y) &=& \frac{3(4x^2 -(1+y^2 -x^2)^2)^2(x^2+y^2-3)^4}{1024x^8y^8}\bigg[\bigg(\ln \frac{|3-(x+y)^2|}{|3-(x-y)^2|} -\frac{4xy}{x^2 +y^2 -3}\bigg)^2 + \pi^2 \Theta(x+y-\sqrt{3})\bigg].\hspace{0.55cm}
\eea
\end{widetext}
The observationally relevant parameter in eqn.(\ref{Omegagw}) represents the fractional energy density residing in GWs and is connected with the power spectrum of the tensor modes, where the brackets indicate the oscillation average  \cite{Kohri:2018awv, Baumann:2007zm}. Eqn.(\ref{Omegaf}) refers to the fraction of the GW energy density as observed today \cite{NANOGrav:2023hvm} and includes the current fractional energy density of radiation $\Omega_{r,0}$ , which is denoted using $``0"$, and $g_{*}(T_{\rm rad})$ refers to the relativistic degree of freedom existing at the temperature $T_{\rm rad}$ during re-entry of the perturbations into the radiation dominated (RD) era. The double overline in eqn.(\ref{Omegagw}) is used to maintain consistency in the notation with the OLRRSPS, employed to derive the tensor power spectrum using the eqn.(\ref{finalps}). The quantity in eqn.(\ref{kernel}) is the integration kernel from the RD era \cite{Kohri:2018awv}. Finally, the frequency and wavenumber follows the relation $f=\small1.6\times 10^{-9}\times (k/10^{6}{\rm Mpc}^{-1}){\rm Hz }.$ 

Fig.({\ref{finalGW}}) describes the SIGW spectra resulting from the use of the OLRRSPS, with $c_s$ values: $c_s=1$ and $c_s=1.17$. For both spectra, we have found in the present context that there exists a resonating behaviour at frequency $f\sim{\cal O}(10^{-7})$ with the experimentally observed signal of the NANOGrav $15$ data having amplitude $\Omega_{\rm GW}h^{2}\sim{\cal O}(10^{-4})$. When plotted for large frequency values $f\sim{\cal O}(10^{-4}{\rm Hz}-10^{4}{\rm Hz})$, the spectra include regions where peaks lie within the sensitivities of the existing and proposed terrestrial and space-based experiments, highlighted accordingly in the figure's background. These peaks originate from enhancements due to MSTs while entering into the several USR (USR$_{n}$) phases. The peak amplitude in the spectra lies within the interval, $\Omega_{\rm GW}h^{2}\sim{\cal O}(10^{-4}-10^{-6})$, for the value of the parameter, $1 \leq c_{s} \leq 1.17$. After the peak values are achieved, the spectrum falls sharply giving a dip-like feature that has an amplitude within the interval, $\Omega_{\rm GW}h^{2}\sim{\cal O}(10^{-12}-10^{-10})$, for the similar allowed interval of the parameter. $1 \leq c_{s} \leq 1.17$. 
During the last SR$_{7}$ phase, the spectra after the enhancement exhibit a sharp fall at $f\sim{\cal O}(10^{12}{\rm Hz})$ which is a result of the final DRG resummation constraint before the end of inflation. It generates sufficient enhancements that correspond to signals for the production of PBHs in the mass range of, $M_{\rm PBH}\sim{\cal O}(10^{-31}M_{\odot}-10^{4}M_{\odot})$, and the necessary region is also highlighted in the form of a band in between the two spectra. The SIGW spectra end for the frequency where inflation successfully ends from the OLRRSPS.

In this letter, our objective was two-fold, the first being the successful generation of a range of PBH masses, thereby evading the existing No-go theorem for heavy mass PBHs. The Second was to have a detailed discussion on the PBH formation mechanism in the presence of renormalization followed by DRG resummation of the OLCSPS and conclude with the generation of the SIGW spectrum. By introducing MSTs in the EFT framework, we found that to have successful inflation with $\Delta{\cal N}_{\rm Total}\sim{\cal O}(60-70)$ requires the use of $6$ short-lived USR phases with $\Delta{\cal N}_{\rm USR_{n}}\sim{\cal O}(2)$, followed by another $6$ SR phases with $\Delta{\cal N}_{\rm SR_{n+1}}\sim{\cal O}(7)$, where $n=1,2,\cdots,6$. The mentioned phases were combined such that the perturbative approximations are respected and produce significant enhancements in the OLRRSPS. With the present MSTs setup, we were able to show the successful generation of PBH masses within the range of $M_{\rm PBH}\sim{\cal O}(10^{-31}M_{\odot}-10^{4}M_{\odot})$. The dynamical nature of $c_{s}$ is frequently utilized in our analysis by taking its value from $1 \leq c_{s} \leq 1.17$. We discuss the case of $c_s=1.17$ with great care as it provides us with the most suitable scenario for PBH formation. We further analyze the extremes of the mass interval - $M_{\rm PBH}\sim{\cal O}(10^{-31}M_{\odot}-10^{4}M_{\odot})$ to identify the range of PBH masses giving a physically significant fractional abundance of $0.001 \leq f_{\rm PBH} \leq 0.5$. Finally, we perform analysis for the generation of SIGWs from the present MSTs setup. From our results, the final SIGW spectrum shows a resonating feature at frequency $f\sim{\cal O}(10^{-9}{\rm Hz})$, which agrees with the NANOGrav $15$ year data and also shows signals within the observed sensitivities of the previously mentioned gravitational experiments. The enhancements in the SIGWs across the range of frequencies, $f\sim{\cal O}(10^{-17}{\rm Hz}-10^{12}{\rm Hz})$, is uniform in the peak amplitude, $\Omega_{\rm GW}h^{2}\sim{\cal O}(10^{-6}-10^{-4})$, and in the dip amplitude, $\Omega_{\rm GW}h^{2}\sim{\cal O}(10^{-12}-10^{-10})$, for the values in interval, $1 \leq c_{s} \leq 1.17$, which also resembles the feature obtained previously for the OLRRSPS. Therefore this letter presents itself as conclusive evidence that a \textit{Strong No-go theorem for the generation of large mass primordial black holes is evaded and successful production of SIGWs is justified from the NANOGrav 15 data as well as with the other established observatories, all with the help of multiple sharp transitions within the framework of EFT}.

\subsection*{Acknowledgements}
SC would like to thank Md. Sami and Sudhakar Panda for useful discussions. SC would like to express gratitude for the work-friendly environment of The Thanu Padmanabhan Centre for Cosmology and Science Popularization (CCSP), SGT University, Gurugram, which provided vast support for this research. Finally, we would like to acknowledge our debt to individuals from various parts of the world for their generous and unwavering support for research in the natural sciences.

\bibliography{references2}
\bibliographystyle{utphys}

\end{document}